\documentclass[prl,twocolumn,superscriptaddress,showpacs,floatfix]{revtex4}\usepackage{amssymb,epsfig}
\def\apj{{Astrophys. J. }}
\def\apjl{{Astrophys. J. Lett. }}
\def\aap{{Astron. \& Astrophys. }}
\def\mnras{{Monthly Not. R. Astron. Soc. }}
\def\prl{{Phys. Rev. Lett. }}

\newcommand{\rem}[1]{ }
\newcommand{\beq}{\begin{equation}}
\newcommand{\eeq}{\end{equation}}
\newcommand{\bea}{\begin{eqnarray}}
\newcommand{\eea}{\end{eqnarray}}

\begin{document}

\title{Cluster magnetic fields from large-scale-structure and 
galaxy-cluster shocks}

\author{Mikhail V. Medvedev} 
\affiliation{Department of Physics and Astronomy, 
University of Kansas, KS 66045}

\author{Luis O. Silva}
\affiliation{GoLP/Centro de Fisica de Plasmas, 
Instituto Superior T\'ecnico, 1049-001 Lisboa, Portugal}

\author{Marc Kamionkowski}
\affiliation{California Institute of Technology, 
Mail Code 130-33, Pasadena, CA 91125}

\begin{abstract}
The origin of the micro-Gauss magnetic fields in galaxy clusters is 
one of the outstanding problem of modern cosmology. We have
performed three-dimensional particle-in-cell simulations of 
the nonrelativistic Weibel instability in an electron-proton plasma,
in conditions typical of cosmological shocks.  These simulations
indicate that cluster fields could have been produced by
shocks propagating through the intergalactic medium during the
formation of large-scale structure or by shocks within the cluster.  
The strengths of the shock-generated fields range from tens of nano-Gauss 
in the intercluster medium to a few micro-Gauss inside galaxy clusters. 
\end{abstract}


\maketitle

The origin of the micro-Gauss magnetic fields observed in galaxy
clusters \cite{Vikhlinin+01,Taylor+02,Vogt+03}
poses one of the most intriguing problems in modern cosmology.
The most common explanation invokes the amplification of seed or
primordial fields by hydrodynamic turbulence that have been
excited during the processes of large-scale-structure (LSS) formation. 
Although there are viable astrophysical mechanisms that can
generate seed fields with $B\sim10^{-16}$~Gauss or 
weaker \cite{Harrison70,Sicotte97,Biermann50,Gnedin+00}, recent
cosmological simulations \cite{Bruggen+05} show that structure
formation can amplify the field by no more than a factor of $10^3$.
Hence, in order to explain the observed intergalactic
field, one needs seed fields as strong as
$B\sim10^{-9}$~Gauss.  Scenarios with galactic winds and
quasar-driven outflows \cite{Miranda+98,Furlanetto+01} can
provide fields of such strength, but they are rather localized.
It is thus not clear whether they can explain entirely the
origin of the intergalactic fields in galaxy clusters.

Here we show that magnetic fields can be
produced by collisionless shocks in galaxy clusters and in the
intercluster medium  (ICM) during LSS formation.
Cosmological $N$-body and hydrodynamic simulations of 
LSS formation \cite{Miniati+00,Ryu+03} have shown
that shocks with Mach numbers up to $M\sim100$ are ubiquitous 
on scales of few to few tens of megaparsecs.
Theoretical analysis of non-magnetized collisionless
shocks indicates that they can generate sub-equipartition
fields \cite{Moiseev+Sagdeev63,Medvedev+Loeb99,Schlickeizer+Shukla03}.
We verify this prediction with state-of-the-art numerical 
simulations. We present here three-dimensional (3D) 
particle-in-cell (PIC) simulations of the {\it nonrelativistic} 
(with $v=0.1c$) Weibel instability \cite{Weibel59,Fried59}
in an {\it electron-proton} (with $m_p/m_e=100$) plasma, 
thus guaranteeing a clear separation of the relevant time scales.  
These simulations are computationally expensive and 
represent a significant advance over previous 
studies, which simulated relativistic shocks $(v\sim c$) in an
electron-positron or low-mass-ratio electron-ion plasmas ($m_i/m_e\le16$) 
\cite{Silva+03,Nishikawa+03,Frederiksen+04}.
Note that a recently discussed possibility that cluster shocks may 
produce the magnetic fields seen in galaxy clusters \cite{Fujita+05} 
was based on the assumption 
that the results of relativistic simulations will also apply in
the nonrelativistic regime. 
Our nonrelativistic simulations fully
confirm theoretical predictions and indicate that LSS shocks can
produce magnetic fields of strengths of tens of nano-Gauss to
few micro-Gauss in the intergalactic medium (IGM) and ICM,
respectively.


The mechanism of the field generation at shocks is rather simple
\cite{Medvedev+Loeb99,Moiseev+Sagdeev63}. As a shock propagates into an 
ambient medium, it reflects (or scatters) a fraction of the incoming 
(in the shock frame) particles back into the upstream region
which then
form counter-propagating streams. Both groups of particles (ICM/IGM and 
reflected particles) have bulk velocities of order 
the shock velocity $v_{\rm sh}$; they can also have some 
thermal spread. Both protons and electrons form the streams, so both 
species participate in the instability \footnote{Unlike the 
ultrarelativistic case, the growth rate of the electrostatic (Langmuir) 
instability may be greater than that of the Weibel instability. The 
electrostatic fields are predominantly longitudinal and scatter the
particles in $z$-direction, not over the pitch angle. The overall
anisotropy of the streams and, hence, the dynamics of the Weibel 
instability are not substantially affected. Therefore, we do not 
consider the Langmuir instability in this paper.
}.
One can consider each charged particle in these streams as an elementary 
current.  Since like currents attract each other, it is energetically 
favorable for the elementary currents to merge into larger current filaments.
This process is inhibited at scales smaller than the plasma skin depth,
$\sim c/\omega_p$ ($\omega_p$ is the plasma frequency), by strong electrostatic
repulsion of like charges. In contrast, on large scales, the currents are
quasi-neutral because of Debye shielding in a plasma. Hence, the
filaments and associated magnetic fields grow rapidly.
The process stops when most of the particles become trapped in the 
produced fields and can no longer amplify the field. 
This happens when the particle Larmor radius 
$\rho_L=v_{\perp B}/\omega_c$ ($v_{\perp B}$ is the particle velocity 
component transverse to the local magnetic field, and $\omega_c=eB/mc$ is 
the cyclotron frequency) becomes comparable to (or less than) 
the characteristic correlation scale $\lambda_B$ of the field,
$\rho_L/\lambda_B\sim1$.
At this time, the particle distribution is effectively
isotropized, and so $v_{\rm thermal}\sim v_{\perp B}\sim v_{\rm sh}$. 

The anisotropy of the particle distribution near a shock can be
parameterized as,
\beq
     A=({\epsilon_\|-\epsilon_\bot})/{\epsilon_{\rm tot}}
     \simeq({M^2-1})/({M^2+1}),
\eeq
where $\epsilon_\|\propto v_{\rm sh}^2$ 
is the energy of the particle along the shock propagation direction;
$\epsilon_\bot\propto v_{\rm thermal}^2\simeq c_s^2$ 
is the thermal energy in the plane of the shock;
$\epsilon_{\rm tot}=\epsilon_\|+\epsilon_\bot$ is the total energy;
$c_s$ is the sound speed upstream; and the Mach number of the shock is
$M=v_{\rm sh}/c_s$. For strong shocks, $M\gg1$, the anisotropy 
parameter is close to unity, $A\sim1$. At a shock, the bulk velocities 
of the electron and proton components are both comparable to the shock
velocity. Hence, the protons dominate over the electrons in the
overall energy budget, and the magnetic field generated by the
electrons is negligible compared with that generated by the
protons. The growth rate and the wavenumber of the fastest growing mode 
(which, in fact, sets the spatial correlation scale of the produced field) are
$\gamma_B=A\,\omega_{p,p} (v_{\rm sh}/c)$ and $k_B=A\,\omega_{p,p}/c$,
where $\omega_{p,p}=\left({4\pi e^2 n_p}/{m_p}\right)^{1/2}
\approx1.32\times10^3 n_p^{1/2}\textrm{  s}^{-1}$
is the proton plasma frequency, and $n_p$ and $m_p$ are the number density 
and the mass of the protons, respectively. (We use cgs units throughout,
unless stated otherwise.) The kinetic calculation of the growth rate and 
the instability threshold have been examined elsewhere \cite{new}.
Order-of-magnitude 
estimates of the  magnetic-field $e$-folding time and the field correlation 
length at strong shocks ($M\gg1$) are readily obtained as
\begin{eqnarray}
\tau_B&\sim&1/\gamma_B\simeq2\times10^{2}\; 
v_{{\rm sh},7}^{-1}n_{\rm ICM,-4}^{-1/2} \textrm{ s},
\\
\lambda_B&\sim&2\pi/k_B\simeq10^{10}\; n_{\rm ICM,-4}^{-1/2} \textrm{ cm},
\end{eqnarray}
for a typical ICM proton density of $n_{\rm ICM}\sim10^{-4}$~cm$^{-3}$ and a
typical shock velocity $v_{\rm sh}\sim10^7$~cm~s$^{-1}$; as usual,
we denote $n_{\rm ICM,-4}=n_{\rm ICM}/(10^{-4}\textrm{ cm}^{-3})$ and 
$v_{\rm sh,7}=v_{\rm sh}/(10^7 \textrm{ cm s}^{-1})$. Since it takes
$N\sim\textrm{few}\times10$ $e$-foldings to produce strong fields, 
we can readily estimate the thickness of a region of the field growth as
$\Delta\sim N\,\tau_B\,v_{\rm sh}\sim N\,\lambda_B$.

The saturation level of the magnetic field is estimated from 
$\lambda_B\sim \rho_L=v_{\rm sh}/\omega_{c,p}$, where 
$\omega_{c,p}={eB}/{m_p c}\approx9.58\times10^3\,B\,\textrm{  s}^{-1}$
is the proton cyclotron frequency. In a multiple-species plasma,
however, saturation occurs at equipartition with the lightest 
species \cite{Wiersma+04}. To incorporate this, we introduce an efficiency 
factor $\eta$, which in the electron-proton plasma is of order $m_e/m_p$.
Finally,
\beq 
\epsilon_B=\frac{B^2/8\pi}{m_p n_p v_{\rm sh}^2/2} 
\simeq\frac{B^2}{8\pi p_{\rm sh}}\simeq A^2\eta\sim 10^{-3},
\eeq 
where $p_{\rm sh}$ is the gas pressure behind the shock, and
the last estimate is for strong shocks, $A\sim 1$.

Although there is no doubt that magnetic fields
are generated at shocks through the Weibel instability, it is
not clear whether they survive
sufficiently far downstream to produce longstanding magnetic fields.
The concern arises from
the fact that the wavelength of the fastest-growing mode in the
linear Weibel-instability analysis is very small,
$\lambda_B\sim2\pi c/\omega_{p,p}\simeq10^{10} \textrm{ cm}$
for a typical ICM particle density of $n\sim10^{-4}$~cm$^{-3}$. Therefore,
it is possible that the extremely short spatial
scales---i.e., sharp field gradients---can be rapidly destroyed by 
dissipation on a plasma time scale of $\tau_B\sim10^2$~s. Should this happen, 
the fields would occupy only a very narrow region near the shock front
and, thus, would not result in long-lived cluster fields.
Recently, it has been shown \cite{Medvedev+05}, both theoretically 
and using relativistic PIC simulations that the correlation length 
of the magnetic field (and, hence, its gradient scale) grows rapidly 
with time, thus drastically reducing diffusive (Ohmic) dissipation. 
These results suggest that the magnetic fields produced should
survive on cosmological times.


In general, it is far from clear that nonrelativistic shocks can generate 
fields in the way relativistic shocks do. In order to test this, we 
have performed a set of 3D and 2D PIC 
simulations \cite{dawson,langdon} using a state-of-the-art,
massively-parallel, electromagnetic, fully-relativistic, 3D PIC
code OSIRIS 2.0 \cite{Fonseca+02}.
In our PIC simulations, 
the initial conditions are taken to be two streams of electrons
and ions moving with relative bulk velocity $v_{\rm sh}$, which
in our simulations we take to be $0.1\, c$.  The four ``species'' of
particles (the upstream and downstream electrons and ions) is
then each assigned a Maxwell-Boltzmann distribution of
velocities about the bulk velocity.  Our ions are ``light
protons,'' positively-charged particles with mass
$m_\mathrm{ions} = 100 \, m_{e}$, a mass ratio large
enough to guarantee that the electron and ion time scales are
clearly separated. All of our
simulations have volumes of $128^3$ cells, although the cell
sizes differ.

We first ran four shorter-duration test simulations, three of
them 3D and one 2D.  In the first 3D run, the box size was
$(25.6~c/\omega_{p,e})^3$ (where $\omega_{p,e}$ is the electron
plasma frequency), and there were 4 particles/species/cell.  The
second simulation differed from the first only in that the box
size was $(12.8~c/\omega_{p,e})^3$, and the third simulation
differed from the first only in that it had 8
particles/species/cell.  These test simulations showed that
neither the box size, nor the number of particles per cell
affect the results within the available computational resources.  
We also ran a 2D simulation with
$1280^2$ cells, i.e, with a box of size $(256.0~c/\omega_{p,e})^2$, with
periodic boundary conditions and 9 particles/cell/species in
order to examine the dynamics of a similar system in 
the plane transverse to the bulk motion of the shocked plasma 
(the configuration is identical to that in simulations of
Ref. \cite{Medvedev+05}). Comparison between the 3D simulations 
and the 2D simulations do not show significant differences 
(e.g., in terms of the saturation level of the magnetic field), 
only revealing the limitations of the 2D simulation  
(the Weibel instability is stronger in 2D), 
and confirming that the transverse dimensions of the 
3D simulation box are not strongly affecting the field 
dynamics on the time scales analyzed here.

We then ran very long 3D simulations of colliding plasma slabs,
for four sets of plasma parameters.  One of the plasma slabs
describes a shocked high--Mach-number plasma ($M = 20$,
$v_\mathrm{th,e\, shock}/c = 0.05$, with either
$v_\mathrm{th,i\, shock}/c = 0.005$ or 
$v_\mathrm{th,i\, shock}/c = 0.0$) with bulk motion 
along the $x_1$ direction. The different ion thermal 
velocities correspond to the two extreme cases of 
a strongly turbulent shock \cite{tidman}, where electrons 
and ions in the shocked plasma are thermalized by plasma 
turbulence and a laminar 
shock \cite{sagdeev,forslund2}.  The second plasma 
slab describes the IGM/ICM plasma with either cold electrons, 
$T_e=0$, or hot electrons, $T_e=T_i \simeq 10  \, \mathrm{eV}$, 
such that $v_\mathrm{th,e\, IGM}/c = 0.05$, 
$v_\mathrm{th,i\, IGM}/c = 0.005$.  All four of these
simulations used a box of size $(25.6~c/\omega_{p,e})^3$.
We found that the different physical parameters did not reveal 
any significant differences in the evolution of $\epsilon_B$.
Our present choice of simulation parameters is strongly 
limited by the time scales involved in the mechanisms
described here, and it aims to illustrate the key features 
of the magnetic-field generation in conditions relevant 
for nonrelativistic collisionless shocks.

\begin{figure}[t!]
\epsfig{file=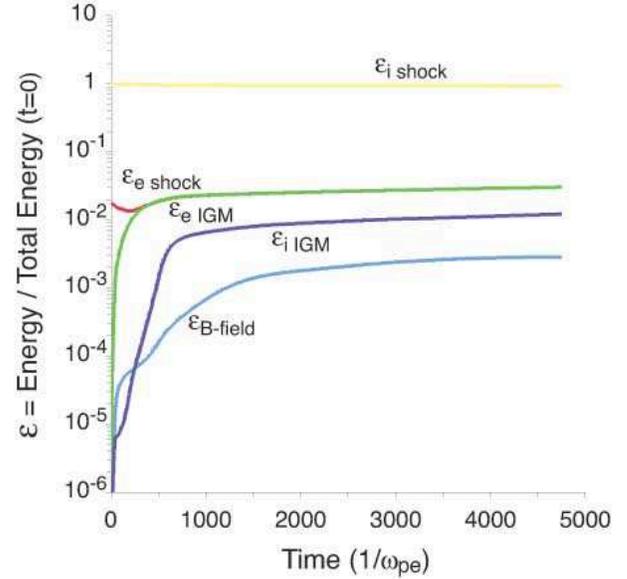, width=86mm}
\caption{
The evolution of the magnetic field energy normalized by the 
total initial kinetic energy, $\epsilon_B$, is shown with the light blue line. 
The energy in the magnetic field is predominantly 
associated with the field components parallel to the shock plane.
For comparison, the similarly normalized energies for the 
four particle species are also shown.}
\label{fig:1}
\end{figure}

In fact, when examining the temporal evolution of 
$\epsilon_B$ measured in the 3D simulations
(Figure \ref{fig:1}), we observe the key role played by 
the ions, with most of the magnetic-field energy
generated by the Weibel instability originating in the 
shocked ions. All the runs revealed 
$\epsilon_B \simeq 10^{-3}$. Note that the Weibel-field growth 
of the ions saturates at lower relative $\epsilon_B$  
than for electrons
$\epsilon_{B,i}\sim(m_e/m_i)^{1/2}\epsilon_{B,e}$,
where for species $s=i,e$, 
$\epsilon_{B,s}=({B^2/8\pi})({m_s n_s v_{\rm sh}^2/2})^{-1}$. 
A strong thermalization between the electrons in the two 
slabs is achieved very early in time via the electron Weibel 
instability, but ion thermalization is not observed 
in our simulations. Other instabilities with longer
time scales (e.g., the ion acoustic instability)
will be responsible for this. These instabilities
are not observed in our simulations since the 
simulation box is not large enough.

\begin{figure}[t!]
\epsfig{file=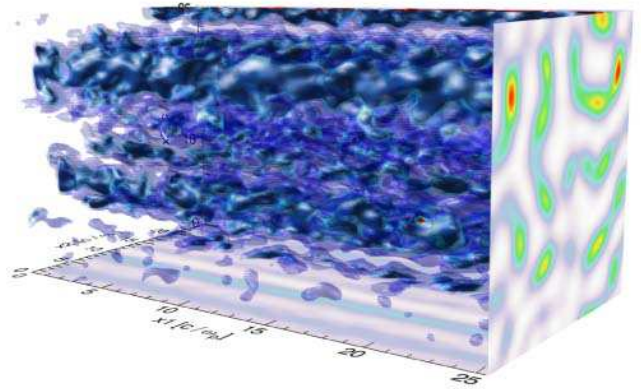, width=86mm}
\caption{
Magnetic field energy density at $t=2000/\omega_{p,e}$ 
($\omega_{p,e}$ is the electron plasma frequency).
The blue iso-surfaces correspond to a value of 
$\epsilon_B \simeq 8 \times 10^{-3}$.
The projection in the $x_2$-$x_3$ plane (the shock plane) is the 
value of $\epsilon_B$ averaged along $x_1$ (the shock propagation direction) 
with red color corresponding to a peak value of 
$\epsilon_B\simeq 6\times 10^{-2}$. The color scale in the
projection plane is linear.}
\label{fig:2}
\end{figure}

The time scale for energy transfer between the ions and 
the magnetic field is the time scale for the Weibel instability 
of the ions, longer than the electron Weibel instability 
by a factor of $(m_i/m_e)^{1/2}$, thus making its observation 
in numerical simulations very time consuming. The structure 
of the generated magnetic field depicted in Figure \ref{fig:2} 
shows the typical configuration of a Weibel-driven field,
in 3D and in the 2D plane transverse to the bulk motion of the shocked 
plasma, surrounding the self-generated current filaments, 
which are already evolving to longer wavelengths.


We have demonstrated that magnetic fields are produced at nonrelativistic  
collisionless shocks and their strengths are comparable to that
observed in clusters.  It is thus natural to explain the
observed fields by the Weibel instability.  If so, then the
magnetization of clusters should begin around the reionization
epoch, at redshifts of $z\sim10-20$, when the gas becomes highly
ionized and particle collisions become rare and
inefficient. Our studies reveal that the magnetic field grows to
an energy density of roughly a tenth of a percent of
the initial kinetic-energy density, and hence constitutes a
similar fraction, $\epsilon_B\sim 10^{-3}$, of the thermal
energy density of the shocked gas. The actual number depends on
complicated nonlinear dynamics of the currents in the downstream
region. This value of the equipartition parameter corresponds to
a magnetic-field strength of order
\beq
     B\sim10^{-8}\;\epsilon_{B,-3}^{1/2}\, v_{{\rm sh},7}\,
     n_{{\rm ICM},-4}^{1/2} \textrm{  Gauss}.
\eeq
These values correspond to tens of nano-Gauss in the ICM
and a few micro-Gauss inside galaxy clusters, the latter in
excellent agreement with observations.

The simulations presented here model a strong shock with Mach 
number, $M=20$. Can a weak, $M\sim1$, shock generate fields as well?
3D simulations of weak shocks are hardly possible at present. However, 
theoretical analysis of the Weibel instability shows 
that fields are generated when the shock velocity is larger than the 
thermal velocity of ICM/IGM particles by a factor of two or more; that 
is, when $M\gtrsim2$. The exact number depends on the actual particle
distribution at the shock.

LSS shocks can be observed via (i) synchrotron
emission by the shock-accelerated electrons in the {\it in situ}
generated magnetic fields; (ii) inverse-Compton scattering of
cosmic microwave background photons by the shock-accelerated
electrons; (iii) Sunyaev-Zeldovich effect on thermal electrons
in the shocked medium downstream; and/or (iv) an abrupt change in the
Faraday rotation measure across the shock. The shock front appears to be
very thin and will likely be unresolved.  Since no sign of
proton thermalization is seen by the end of the simulations,
$t\sim500\,\omega_{p,p}$, we can put a constraint on the shock
thickness,
\beq
     \Delta_{\rm sh} > 10^{13}\; n_{{\rm ICM},-4}^{-1/2} \textrm{  cm}.
\eeq

Our present analysis does not consider the evolution 
of the fields on cosmological time scales. Numerical simulations 
of this type are hardly possible within the next few years. Theoretical 
considerations suggest that inverse cascade should result in the rapid 
transfer of magnetic energy from small (shock) to large (cosmological) scales, 
thus leading to the long-term survival of the fields \cite{Medvedev+05}.
An alternative possibility, that other instabilities \cite{Milos+Nakar05} 
present in plasma could destroy the field entirely, 
is unlikely because the time scales involved are long enough
for dynamos driven by turbulence and sheared motions of gas in clusters
to further amplify and preserve the shock-generated magnetic fields.
A detailed study of these issues is highly important, yet extremely 
difficult and should be addressed in the future.

\begin{acknowledgments}
LOS acknowledges the help of Michael Marti in performing the simulations, 
M.~Tzonfres, M.~Marti, Profs. Ricardo Fonseca and Warren Mori for 
discussions.  MK acknowledges
useful discussions with X.~Chen, E.~Nakar, and M.~Milosavljevic.
The simulations were performed in the eXpp cluster at IST,
Lisbon. The work of MVM has been supported by
DoE grant DE-FG02-04ER54790, NASA grant NNG-04GM41G, and the GRF fund.
The work of LOS was partially supported by FCT (Portugal)
through grants PDCT/FP/FAT/50190/2003 and POCI/FIS/55905/2004.
MK was supported by DoE DE-FG03-92-ER40701 and NASA NNG05GF69G.
\end{acknowledgments}

~\\ 

~\\
\newpage

\end{document}